# Magnetised quark nuggets in the atmosphere

T. Sloan[1,*], J. Pace VanDevender[2], Tracianne B. Neilsen[3], Robert L. Baskin[4], Gabriel Fronk[3], Criss Swaim[5], Rinat Zakirov[2], and Haydn Jones[2]

[1]Department of Physics, Lancaster University, Lancaster, LA1 4YB, UK
[2]VanDevender Enterprises LLC, 7604 Lamplighter LN NE, Albuquerque, NM 87109 USA
[3]Department of Physics, Brigham Young University, Provo, UT, USA
[4] Gestalt GeoResearch, 819 Springwood Drive, North Salt Lake City, Utah, USA
[5]The Pineridge Group, LLC, 6 Delwood Circle, Durango, CO 81301 USA

*t.sloan@lancaster.ac.uk

 Abstract

A search for magnetised quark nuggets (MQN) is reported using acoustic signals from hydrophones placed in the Great Salt Lake (GSL) in the USA. No events satisfying the expected signature were seen. This observation allows limits to be set on the flux of MQNs penetrating the Earth's atmosphere and depositing energy in the GSL. The expected signature of the events was derived from pressure pulses caused by high-explosive cords between the lake surface and bottom at various locations in the GSL. The limits obtained from this search are compared with those obtained from previous searches and are compared to models for the formation of MQNs.

## Introduction

QNs, (sometimes called nuclearites or strangelets), are a novel form of matter consisting of u, d and s quarks. They were first proposed by Bodmer [1] and Witten [2] who pointed out that an aggregation of such quarks could be stable since there are three Fermi spheres to fill rather than just two for ordinary nuclear matter, which consists of u and d quarks bound within the confines of neutrons and protons. The extra Fermi Sphere could lower the energy state for the quarks binding the higher-mass strange quark, overcoming the tendency for the strange quark to decay, and negating the effects of surface tension seen in nuclei. Since the quarks are not constrained by the need to be held within a nucleon, the density of this state of matter could be higher than the nuclear density. Furthermore, since the lowest energy levels in the Fermi spheres would be filled in the ground state, QNs would be nearly electrically neutral. Hence the repulsive electrostatic forces which decrease the stability of higher mass nuclei would be absent for QNs, and there would not be any limit to their masses. Indeed, Fahri and Jaffe [3] showed that large masses are possible.

QNs could be candidates for the Dark Matter (DM) in the Universe. Furthermore, Liang and Zhitnitsky [4] postulated that they could explain the matter-antimatter asymmetry in the Universe. Hence identification of QN DM could solve two of the major outstanding problems in cosmology.

Tatsumi [5] showed that QNs could also show ferromagnetism through the one-gluon-exchange mechanism. There should then be a large magnetic field on the surface of such QNs with a predicted strength of $B_o \sim 10^{12\pm1}$ T (i.e., 0.1 to 10 Tera Tesla, TT), typical of that at the surface of a neutron or proton.

In this paper the results of a search for MQNs are described. An acoustic technique was employed using hydrophones placed in the Great Salt Lake (GSL) in the USA. In Ref. [6], we presented the design and supporting simulations for the experiment in the GSL. The results of the experiment are reported here.

Previous searches for QNs have been made using underground detectors [7-11]. These experiments would not have been sensitive to MQNs which would range out in the rock overburden. The searches for non-ferromagnetic QNs by Porter *et al.* [12] and Piotrowski, *et al.* [13] using astronomical telescopes would have been sensitive to MQNs but their flux limits were given for non-ferromagnetic QNs. A previous study by our group [14] excluded MQNs with very high surface magnetic fields. The experiment presented here is a search for MQNs with lower surface magnetic fields.

**Properties of Quark Nuggets.**

**(a) Non-ferromagnetic QN properties**

The properties of non-ferromagnetic QNs, discussed by De Rujula and Glashow [15], were used in the published searches with astronomical telescopes for non-ferromagnetic QNs by Porter *et al.* [12] and Piotrowski *et al.* [13].

A brief summary of these properties follows. QNs could have densities somewhat greater than nuclear density ($\sim 2 \times 10^{17}$ kg/m³). They are larger than atomic sizes for masses greater than $10^{-12}$ kg. Their interaction cross section is the geometric area of the QN. Hence, the force on a QN passing through matter equals the specific energy loss (sometimes referred to as the stopping power) and is given by

$$\frac{dE}{dx} = -K\pi R_{qn}^2 \rho v_{qn}^2, \qquad (1)$$

where $R_{qn}$ is the radius of the QN (assumed spherical), $\rho$ is the mass density through which the QN is passing, $v_{qn}$ is the QN velocity, and $K$=2/3 assuming the molecules are in random motion or 4/3 assuming they are stationary. Here we set $K$=1, the difference reflecting the uncertainties in the assumptions. If QNs are constituents of DM, they would be expected to enter the solar system at speeds, $v_{qn} \sim 230$ km/s [16], the velocity of the Solar System in its path through the Galaxy.

From equation (1), the range of travel $R$ of a QN of mass $M_{qn}$ in material of density $\rho$ is

$$R = \int_{v_{qn}}^{v_{min}} \left(\frac{dE}{dx}\right)^{-1} dE = \frac{1}{\rho}\left(\frac{M_{qn}}{\pi}\right)^{1/3}\left(\frac{4}{3}\rho_{qn}\right)^{2/3} \ln\frac{v_{qn}}{v_{min}}, \qquad (2)$$

where $v_{min}$ is the final velocity at which the QN is taken to be at rest. The value of this is of order 1000 m/sec in a solid, a velocity below which the QN has insufficient kinetic energy to break the molecular bonds, but somewhat lower for a gaseous material. However, the range is insensitive to the precise value. Equation (2) shows that non-ferromagnetic QNs of mass more than $10^{-6}$ kg will pass through the Earth without stopping (for $\rho_{qn} = 2 \times 10^{17}$ kg/m³ and $\rho \approx$ 1000 kg/m³).

**(b) Ferromagnetic QN properties**

MQNs have been shown to develop a magnetopause while passing through a medium [6]. A plasma of free electrons and positive ions is created as the electrons are torn off the atoms due to friction and their interaction with the intense magnetic field. The plasma comes into equilibrium at a distance $R_{mp}$ from the MQN at which the plasma pressure balances the pressure exerted by the magnetic field. Assuming that the plasma-facing surface of the magnetopause is approximately spherical, the magnetopause has been shown to have a radius, $R_{mp}$ given by

$$R_{mp} \approx \left( \frac{2B_o^2}{\mu_0 \rho v_{qn}^2} \right)^{\frac{1}{6}} R_{qn} \qquad (3)$$

where $B_0$ is the magnetic field at the equator of the QN and $\mu_0$ is the permeability of free space. Assuming that the cross section is given by the area of the magnetopause, the value of the stopping power is given by substituting $R_{mp}$ for $R_{qn}$ in equation (1).

$$\frac{dE}{dx} = \pi R_{mp}^2 v_{qn}^2 \rho = \pi \left(\frac{2B_0^2}{\mu_0}\right)^{1/3} \rho^{2/3} v_{qn}^{4/3} R_{qn}^2. \qquad (4)$$

To obtain the range (to within a small factor of order unity), this is integrated as in equation (2).

$$R = \int \left(\frac{dE}{dx}\right)^{-1} dE = \alpha (v_{qn}^{2/3} - v_{min}^{2/3}) \quad \text{with } \alpha = \frac{M_{qn}}{\pi (B_0^2 R_{qn}^6 / \mu_0)^{1/3} \rho^{2/3}}.$$

(5)

Since the size of the magnetopause is much greater than the radius of the QN, the stopping power is larger for MQNs than for non-magnetic QNs and, therefore, the range in matter is much shorter.

**Acoustic detection of MQN impacts into the Great Salt Lake, Utah, USA, after atmospheric transit**

**(a) The hydrophone array and its calibration.**

As described in Supplemental Information on Methods: Sensor locations, topology and sound speed of the Great Salt Lake (GSL) Observatory for MQNs, an array of three hydrophones sensors was placed in the GSL at a depth of 2 m (GSL mean depth 4.8 m) for the purpose of detecting MQN impacts. The hydrophones and data acquisition system are described in detail in Ref. [6] and briefly updated with important new information in Methods below. The hydrophones, labelled a1, a2 and a3, were set 300 m apart on the vertices of an equilateral triangle and were each linearly sensitive to pressure variations of up to 112 Pa. As described in Supplemental Information on Methods: Explosive calibrations emulating MQN impacts, each of the three sensors were calibrated using explosive shots from a 5m long PETN primer cord with energy depositions of 130 kJ/m and 260 kJ/m.

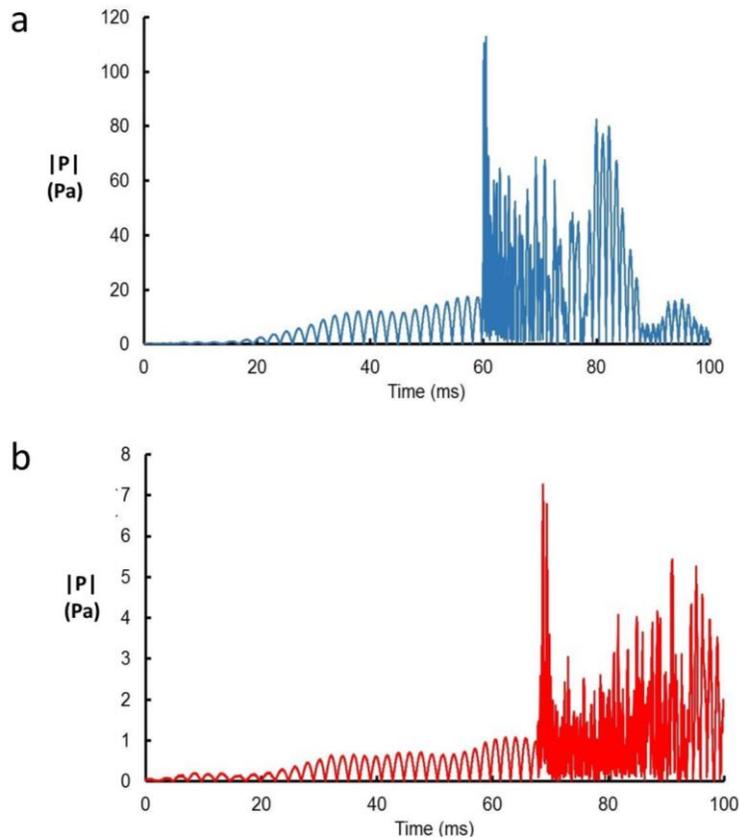

**Figure 1.** Absolute rectified value of the pressure from calibration shots at a sensor a) at 3100 m and b) 4100 m distance from a 5-m length of PETN primer cord depositing 260 kJ/m energy/length. The cord extended vertically from the surface to the lake bottom.

Fig. 1 (and Fig. S4 in the supplementary information) shows typical waveforms resulting from the calibration shots. The pressure waveforms were observed to have a nearly constant-frequency component arriving in the early part of the signal, at frequency ~250 Hz, followed by a higher-frequency, broadband signal. The time duration of the ~ 250-Hz component varies linearly with distance, $d$, as $0.017d$. This 0.017 factor indicates that the group velocity of the ~ 250-Hz part of the waveform was 2.7% greater than that for the higher frequency part.

The shots at near distances did not have time for the ~250-Hz portion of the waveform to reach full amplitude; therefore, each had to be extrapolated to a fixed time (chosen to be 60 ms when the amplitude had reached its plateau value). Fig. 2 shows the peak amplitude in the ~250-Hz part of the waveform at 60 ms after the trigger as a function of distance together with linear fits to the data. The rate of change with distance indicates that the ~250-Hz part of the waveform is attenuated at a rate of $22\pm 3$ dB/km. The summed pressure in the higher frequency part of the waveform was observed to be attenuated at the rate of $26\pm3$ db/km. This decay rate was observed for distances greater than 2.9 km, since shots closer to the sensors saturated the system in the higher frequency portion of the waveform. The difference between the linear fits to the data in Fig. 2 and in the peak levels of the distant shots indicates that the pressure at the sensor varies linearly with instantaneous energy deposition.

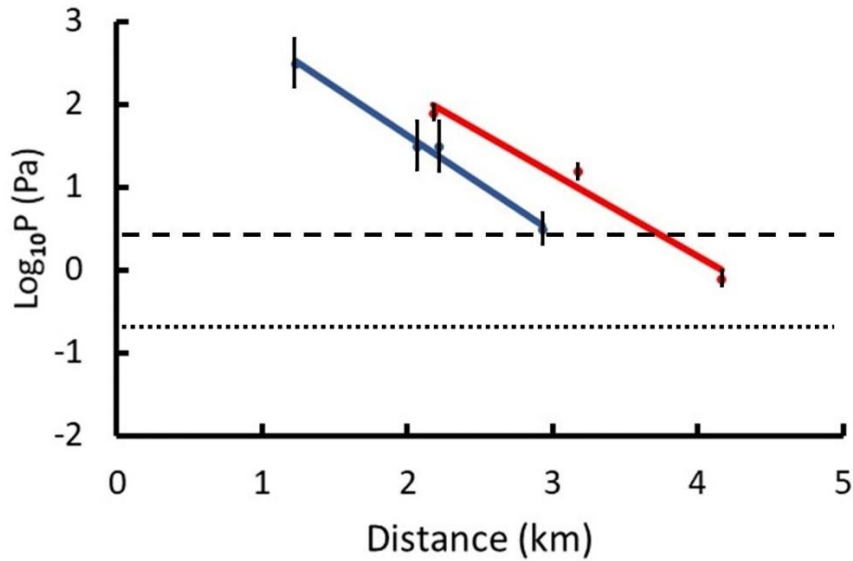

**Figure 2.** Peak pressure at the sensor in the ~250 Hz oscillatory portion versus distance to the source for explosive calibrations using 130 kJ/m (blue) and 260 kJ/m (red) charges, each 5 meters long. The two horizontal lines give the pressure that will trigger recording 80% (dashed line) and 20% (dotted line) of the total observation time, as explained in the text and in Supplemental Information on Methods: Variation in trigger level from weather effects.

As explained in Supplemental Information on Methods: Sensor locations, topology and sound speed of the Great Salt Lake (GSL) Observatory for MQNs, we measured the sound speed $c_s$ as function of depth z. The data show a gradient $\Delta c_s/\Delta z \sim 0.5$ s$^{-1}$, which means acoustic waves in the water are refracted along a circular trajectory with radius of ~ 3.2 km. Consequently, propagation is quite complicated. Simple estimates of pressure versus distance are incorrect. The multiple layers of material in and under the GSL had to be considered to understand propagation.

The geology under the Great Salt Lake is complex with layers of mud, mirabilite, and rock and with a fault line separating the east and west sides of the lake. Seismic exploration and core-sampling have been focused on the east side, but our experiments had to be on the west side to avoid boat traffic.

In spite of the complexity, the acoustic signals from explosive calibrations emulating an MQN impact created the relatively simple waveform shown in Fig. 1. To better understand these signals, computer simulations were carried out using the ORCA package [17]. Three-dimensional effects, variations in the actual depth, and remaining uncertainties in the actual lake-floor geology were approximated by a simulation that assumed horizontal stratification of the water column and sediment layers.

The 5 m of brine with the measured 25% NaCl by weight was modelled with sound speed varying from 1595 m/s (at the top) to 1597 m/s (at the bottom), with a density of 1200 kg/m$^3$, and compressional attenuation of 2.0 dB/m to include absorption and refraction of sound in the lake [18]. The simulated thickness of the layers of mud, mirabilite, and rock were consistent with extrapolations from the measurements [19] on the east side of the lake. The properties of each layer were assumed to be independent of radius and azimuth. Idealized simulations, based on a continual line source, a single receiver, and a lake of uniform depth, showed that acoustic energy was propagated in a waveguide formed by the multiple layers and produced a resonant signal at a frequency of approximately 250 Hz followed by a higher frequency signal burst, as was observed with the calibration shots. The signal properties were found to be very sensitive to the precise thicknesses of the individual layers. A combination of 5 m of brine, 0.75 m of hard mineral (corresponding to the stromatolite layer we found in this part of the lake), 3.0 m of mud, a 0.6 m transition layer into a 1.8 m mirabilite mineral layer above a 5 m rock layer, over bed-rock (to bound the simulation) reproduced the observed group velocity difference between the resonant and higher frequency components as well as the attenuation of the signals observed from the calibration shots (Fig. 2 and supplementary information Fig. S4 and also Fig. S7). The simulation results give us confidence that our approximation of the acoustical properties of the Great Salt Lake is sufficient to model key features observed in the signals from the calibration shots. Thus, since the calibration shots emulate the energy deposition of MQNs, an MQN should have given a signal that would have appeared on the hydrophone recordings.

**(b) Data Taking.**

The experiment was deployed for a total of 146 days, The threshold for triggering the sensors depended on ambient weather conditions. The trigger threshold was reset every hour to three times the maximum amplitude of 10,000 digitized samples taken at 1,000,000 samples per second for a duration of 0.01 seconds. This implies the trigger level was a minimum of 13.5 standard deviations above the random ambient background. Continually resetting the threshold meant that the total observation time depended on signal amplitude. Setting the trigger level at a pressure of 2.5 Pa meant that the apparatus was sensitive for 80% of the time. The total data taking time allowing for all losses of sensitive time was $7.8 \times 10^6$ seconds. The thresholds for reliably capturing the MQN signature are shown in Fig. 2 during >80% and >20% of weather-limited observation time.

Coherently produced triggered events were recorded if the amplitude exceeded this threshold for a duration of more than 3 ms. That duration was set to exclude signals that were obviously incompatible with the signatures from the explosive calibrations. Each hydrophone was triggered autonomously, recording the digitised signal as the mean of ten samples every 10 μs together with a time stamp (from the GPS).

**(c) Data Analysis**

During the portion of the data taking that had all three sensors on line, sensors a1, a2, and a3 were respectively triggered 5,865,531, 1,941,741, and 142,282 times. The calibration shots showed similar behaviour for all 3 hydrophones. The rather different trigger rates associated with the 3 sensors could not be ascribed to ambient conditions but were thought to be generated electrochemically by the corrosive environment of the GSL. Sound transited the distance between sensors in ≈0.2 seconds. To ensure that variations in trigger levels did not exclude recording of coincident signals, a time interval of 0.7 seconds was used as an initial filter for the data. The number of coincident triggers in all 3 sensors within a time of 0.7 seconds was 1110. The signals are mostly of short duration. The 3 longest duration signals are shown in Fig. 3.

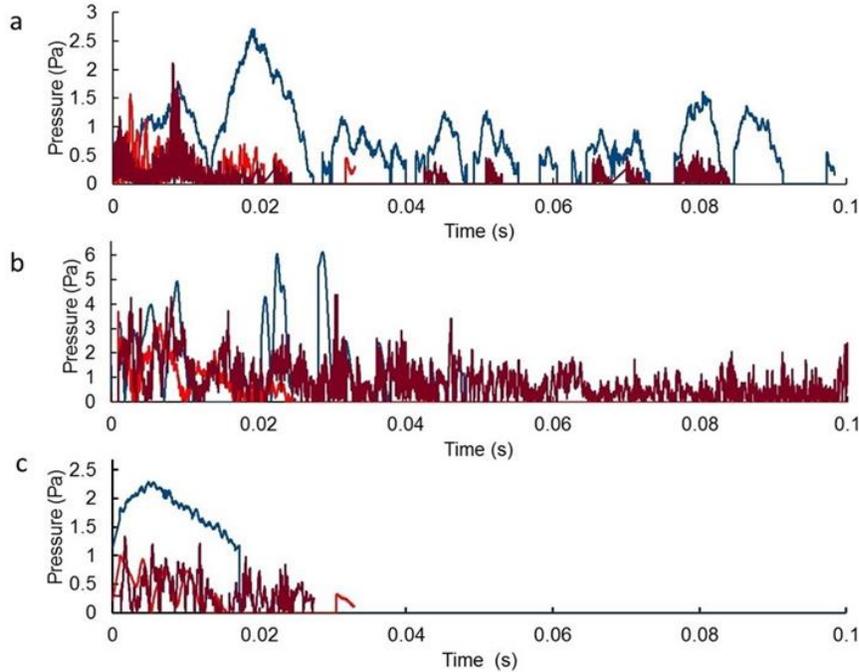

**Figure 3.** Pressure versus time for 3 longest duration signals recorded on (a) 11 May 2019, (b) 2 July 2019, and (c) 14 September 2019 for sensor a1 (blue), a2 (bright red) and a3 (dark red).

It can be seen that these signals do not resemble those from the explosive calibrations (see Fig. 1 and Supplemental Information Fig. S4). Unlike the calibration signals, these waveforms have no constant-frequency part preceding a high-frequency portion and they vary in duration and amplitude for the different sensors. They are composed of sequences of separate pulses separated by times when the pressure is below the trigger threshold of 2.5 Pa. Only one of the hydrophones had a signal above threshold in each case. For these reasons this event could not have been caused by the passage of an MQN and such events were rejected.

**Limits on the Flux of MQNs.**

From the above, we conclude that we did not observe any candidate MQN within the exposure time. This means that, applying Poisson statistics, at 90% confidence level there would have been less than 2.3 possible MQN events within the exposure time. Hence the minimum detectable flux of MQNs, $\Phi$, is given by

$$\Phi = \frac{2.3}{2\pi\varepsilon TA} = \frac{7.4\times10^{-16}}{\varepsilon} \text{ m}^{-2} \text{ s}^{-1} \text{ sr}^{-1}, \tag{6}$$

where $T = 7.8 \times 10^6$ seconds is the exposure time, A is the effective area sampled by the sensor array, and $\varepsilon$ is the acceptance of the apparatus. The sensitive radius was taken to be 4.5 km, the range over which the explosive calibration shots were carried out, giving A=63.6 km² in equation (6).

The acceptance, $\varepsilon$, is computed as follows. The DM is assumed to be composed of discrete MQN particles following the velocity distribution in the model of Read (16). In this model, MQNs have a random velocity $\vec{v_r}$ approximated by a Maxwell-Boltzmann distribution with the most probable velocity of 220 km/s. Superimposed on this random velocity is a streaming velocity $\vec{v_s}$ = 230 km/s caused by the orbit of the Solar System through the Galaxy, currently in the direction of the constellation Lyra. The net velocity of an MQN is, therefore, $\vec{v_f} = \vec{v_r} + \vec{v_s}$. The total flux of MQNs, $\Gamma_{tot}$, heading in the direction of the GSL is given by

$$\Gamma_{tot} = \iiint \frac{\rho_{DM}}{M_{qn}} \frac{\vec{S}\cdot\vec{v_f}}{|S|} P(\vec{v_f}) d^3\vec{v_f} = \iiint \frac{\rho_{DM}}{M_{qn}} \frac{\vec{S}\cdot\vec{v_f}}{|S|} P(\vec{v_r}) d^3\vec{v_r} \frac{d^3\vec{v_f}}{d^3\vec{v_r}}, \tag{7}$$

in which the Jacobian $\frac{d^3\vec{v_f}}{d^3\vec{v_r}} = \frac{v_f^2}{v_r^2} \frac{\partial(v_f,\cos\theta_f,\varphi_f)}{\partial(v_r,\cos\theta_r,\varphi_r)}$, $v_f$ and $M_{qn}$, are the MQN velocity and mass, $\theta_f$ and $\varphi_f$ are the polar and azimuthal angles, $\vec{S}$ is an area vector perpendicular to the GSL, $\cos\alpha = \vec{S}\cdot\vec{v_f}/|v_f|$ allows for the angle between the MQN track and the normal to the GSL, $\rho_{DM} = 7.4 \times 10^{-22}$ kg/m³ is the DM density, $P(v_f,\cos\theta_f,\varphi_f)$ is the probability for the MQN to appear in the phase space volume element $v_f^2 dv_f d\cos\theta_f d\varphi_f$.

This probability can then be expressed as

$$P(v_f,\cos\theta_f,\varphi_f) d^3\vec{v_f} = P(v_r,\cos\theta_r,\varphi_r) d^3\vec{v_r} \frac{d^3\vec{v_f}}{d^3\vec{v_r}}, \tag{8}$$

in which the random Maxwell-Boltzmann probability is given by

$$P(v_r,\cos\theta_r,\varphi_r) dV_r = \left(\frac{1}{\pi v_p^2}\right)^{\frac{3}{2}} v_r^2 e^{-\frac{v_r^2}{v_p^2}} dv_r d\cos\theta_r d\varphi_r. \tag{9}$$

The expression for the detected flux is similar to equation (7) with an extra probability $Q(\vec{v_f})$ inserted to represent the probability that the MQN is detected in the hydrophone array:

$$\Gamma_{det} = \iiint \frac{\rho_{DM}}{M_{qn}} \frac{\vec{S}\cdot\vec{v_f}}{|S|} P(\vec{v_f}) Q(v_f) d^3\vec{v_f} = \iiint \frac{\rho_{DM}}{M_{qn}} \frac{\vec{S}\cdot\vec{v_f}}{|S|} P(\vec{v_r}) Q(\vec{v_r}) d^3\vec{v_r} \frac{d^3\vec{v_f}}{d^3\vec{v_r}}, \tag{10}$$

The acceptance is then the ratio $\Gamma_{det}/\Gamma_{tot}$. The integrals in equations (7) and (10) were computed by Monte Carlo technique.

For each Monte Carlo event, the MQN was tracked through the atmosphere and the pressure, $P$ in Pascals, detected by the hydrophone array computed from the energy $E_{dep}$ in Joules deposited in the GSL as

$$P = \frac{0.453\, E_{dep}}{S} \exp(-2.72\, d) \text{ Pa.} \qquad (11)$$

Here $d$ is the distance (in km) of the simulated hit position to the hydrophone array and $S$ is a safety factor taken to be 2. The quantities in equation (11) were determined from the peak pressures seen in Fig. 1 and the fits to the calibration shot data in Fig. 2 and Fig. S4 in the supplementary information. The factor $S$ is included to allow for the uncertainty in determining the parameters in equation (11). It errs on the side of caution since it lowers the acceptance, increasing the determined flux limit. If the calculated pressure exceeds the threshold of 2.5 Pa the Monte Carlo event is accepted allowing $\Gamma_{det}$ to be determined whilst $\Gamma_{tot}$ comes from all events generated. The acceptances are reasonably insensitive to these values since most of the failure to detect MQN events comes from their stopping in the atmosphere. When they penetrated to the GSL, they usually deposited enough energy to trigger the sensor.

The limits determined from this procedure are shown in Fig. 4 for different values of the surface magnetic field and mass density of the MQN.

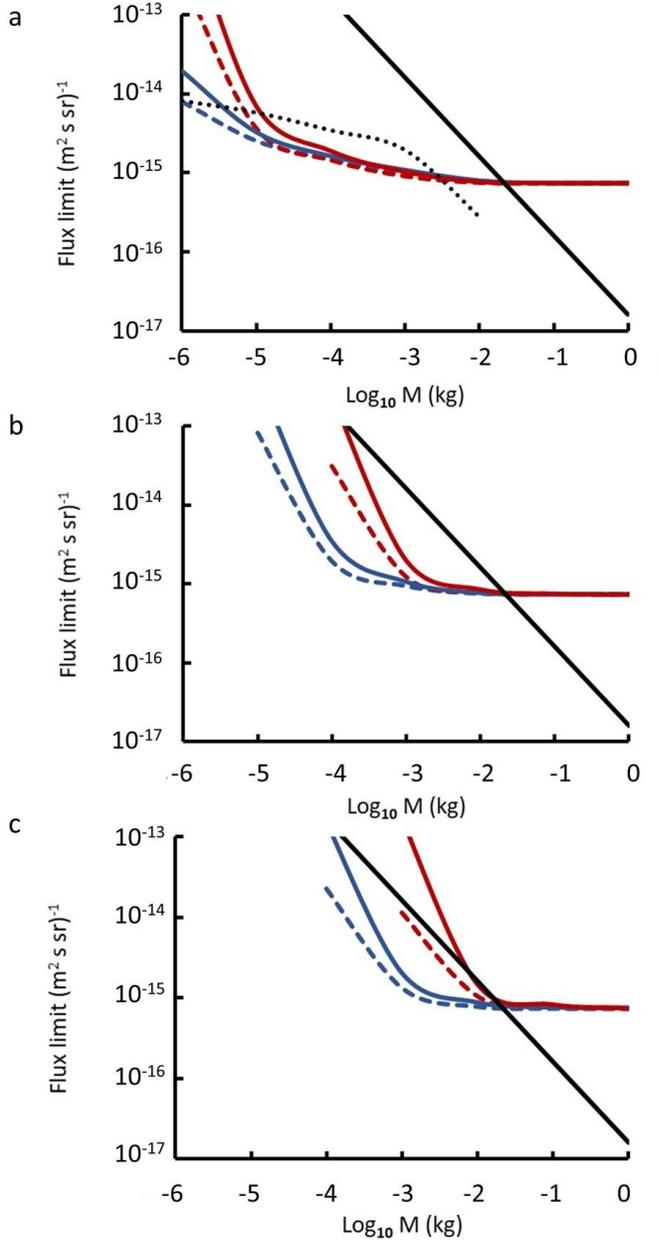

**Figure 4.** Flux limits as a function of MQN mass for 3 different surface magnetic fields $B_o$ of the MQN: a) 0.1 TT, b) 1.0 TT, and c) 3.2 TT. The blue (red) lines are for MQN density of $10^{18}$ ($3 \times 10^{17}$) kg/m$^3$. Two curves for each density show the observationally determined limits with zero streaming velocity (solid) and with a streaming velocity of 230 km/s through the Galaxy (dashed). The solid black line is the theoretically expected flux on the GSL assuming all the DM is made up of MQNs of a fixed mass. The black dotted line is the integrated flux greater than the mass on the abscissa theoretically predicted from the Aggregation Model (see text) for $B_o$=0.1 TT. Curves for higher $B_o$ have flux limits $< 10^{-17}$ m$^2$ s$^{-1}$ sr$^{-1}$ (do not fit on the plots) and are not constrained by the experimental flux limits. Values of $B_0 > 3.2$ TT were excluded as described in Ref. [14].

The acceptance tends to unity at higher masses since such MQNs have higher kinetic energy and do not range out in the atmosphere. In addition, they have larger stopping powers and, hence, give larger signals on passage through the water of the GSL. Lower mass MQNs tend to either range out in the atmosphere (the dominant contributor to the inefficiency) or give smaller signals on passage through the water of the GSL. Hence, they have smaller acceptances.

Acceptance is relatively insensitive to the model as shown in Fig. 4 by the pairs of curves with the streaming velocity set to zero and with the full streaming velocity. However, the acceptance is sensitive to the assumed mass density of the MQN and value of $B_o$ (compare the dashed and solid curves in Fig. 4 for each $B_o$). The reason is that the stopping power is larger for lower MQN density (varying as $\rho_{qn}^{-2/3}$) so that lower density MQNs have shorter ranges in the atmosphere.

**Comparison of the limits with models**
a) **Comparison with the fixed mass model**

In this model, all the DM is assumed to be concentrated in MQNs all with the same fixed mass. The total flux of MQNs hitting the GSL surface determined from equation (7) for our planar detector is then $1.5 \times 10^{-17}/M_{qn}$. This is shown as the close dotted line on Fig. 4. In this case the mass of the MQNs would have to be greater than $10^{-2}$ kg for the expected flux to be too small to have been detected in this experiment. Hence this model is ruled out unless the MQNs have masses greater than $10^{-2}$ kg.

b) **Comparison with the Aggregation Model**

The open dotted curve on the upper panel of Fig. 4 shows the integrated flux for MQNs of mass greater than that shown on the abscissa from the Aggregation Model [14]. The model utilises the ΛCDM model of the Universe starting at T=100 MeV at a time of 65 μs. The DM in the Universe was assumed at this time to consist of a sea of MQNs with A=1 i.e., single u-d-s quark nuggets. The aggregation of these via their long-range magnetic dipole interactions was simulated assuming that two ferromagnetic MQNs would coalesce into a larger MQN if the attractive potential energy between the dipoles is greater than their separation kinetic energy. The simulation showed that the size of the objects grows much faster than the characteristic decay time of the strange quarks (~$10^{-10}$ seconds). Hence, it is quite plausible that MQNs would grow quickly enough for their increasing inter-quark binding energy to prevent such decays. The growth of MQNs depends sensitively on the value of the surface magnetic field, $B_0$. Higher values concentrate the primordial MQNs into higher mass MQNs before the mass distribution freezes out as the universe expands. In consequence, the flux at lower masses falls rapidly with increasing surface magnetic field $B_o$.

Fig. 5 shows the predictions of this model for different values of the surface magnetic field, of the MQN together with the flux limit from this experiment (bottom of the black triangle). We conclude that in the Aggregation Model the surface magnetic field of the MQNs must be greater than 0.12 TT otherwise a signal would have been detected in this experiment.

Limits have been set on the fluxes of non-ferromagnetic quark nuggets (QNs) by the experiments of Porter, *et al.* [12] and Piotrowski, *et al.* [13] using astronomical telescopes. If we assume that the acceptances in these experiments for MQNs is similar to those for QNs their flux limits can also be used to limit the value of $B_o$. The bottom of the blue triangle in Fig. 5 shows their limits with this assumption.

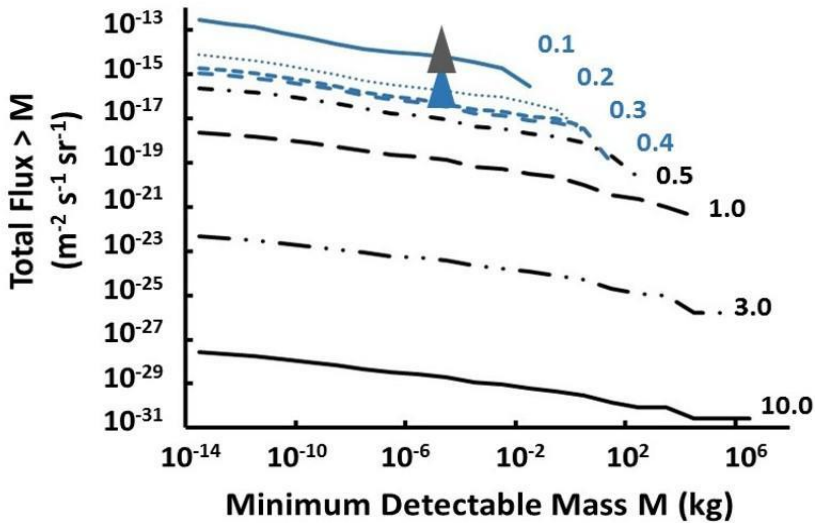

**Figure 5.** Plot of total MQN flux for mass greater than M as a function of $B_o$. The bottom of the grey triangle shows the flux limit for exclusion by the data from the GSL experiment with 90% CL. The bottom of the blue triangle indicates exclusion by the telescope data from Piotrowski, *et al.* and Porter, *et al.* applied to MQNs.

**c) Comparison with other experiments**

There have been previous searches for non-magnetic QNs [20, 21] but we believe that none have addressed the case of MQNs. The high-altitude search by the SLIM collaboration [8] would have been sensitive to MQNs but their published flux limits are orders of magnitude higher than those published here. The deep underground searches by the MICA [10], MACRO [11]. and OHYA [9] collaborations have published limits on the QN flux which are again higher than the limits presented here due to the limited area of the detectors. However, these experiments would only have been sensitive to very high mass MQNs since the lower masses would range out in the rock overburden. The new experiments CTA [22] and JEM-EUSO [23] have the potential to perform searches for QNs and MQNs over a wide area. However, triggers to detect very fast meteor trails will be needed to distinguish solar system meteors from extra-galactic MQNs.

**Discussion**

Other methods for detecting MQNs include measuring non-meteorite impacts on Earth [24, 25] and measuring their radio-frequency emissions after passing through Earth's magnetosphere [26]. Each method has strengths and weaknesses. Tables in Ref. [14] Supplementary Information provide quantitative mass distributions for evaluating a specific detection system. The calculation in this paper assumed the mass density of Dark Matter in the solar system is the same as it is in interstellar space. That may not be true for MQNs since some of them would pass through a portion of the Sun, be slowed to less than escape velocity from the solar system, and be subsequently scattered by the combined planetary gravity to prevent their return to the Sun. These will be trapped in the solar system and enhance the density of Dark Matter. We do not yet know the enhancement factor, so the event rates described above may be pessimistic. If so, low-cost efforts to look for QNs and MQNs are recommended. A special trigger to detect fast meteors in a high-energy gamma-ray cosmic-ray detector could be used to prove or disprove, or place tighter limits on the existence of these objects.

**Conclusion**

A search for ferromagnetic quark nuggets (MQNs) has been undertaken. No significant signal was observed and limits on the flux of these objects hitting the Earth have been set. In the fixed mass model masses of MQNs up to $10^{-2}$ kg are excluded. In the Aggregation Model of Ref. [14] magnetic fields of less than 0.12 TT are excluded by the data presented here.

In order to detect these objects, if they exist, experiments covering larger fields of view with longer exposure times will be required. The atmosphere is a barrier for MQNs since low mass MQNs stop before reaching the Earth's surface. The Aggregation Model predicts that most of the dark matter density will be concentrated in the high mass region whereas most of the flux will be at low masses. Hence searches in the upper atmosphere or in space will be advantageous. An experiment sensitive to fast high-altitude meteors could detect MQNs since most meteors travel with the velocity of the Earth around the solar system whilst QNs and MQNs which are of Galactic origin will have velocities about 10 times larger.

It would be good to have a dedicated experimental program to search for these objects, given that either of them may carry the solution for two of the outstanding problems in cosmology, namely the nature of the dark matter in the Universe and the matter-antimatter asymmetry in the Universe.

**Methods**

<u>Hydrophones and Data Acquisition Systems</u>
The hydrophones and data acquisition system are described in detail in Ref. [6]. Each data acquisition system used a N210 USRP (Universal Software Radio Peripheral) manufactured by National Electronics and used the gnuradio open-source UHD (Universal Hardware Driver) to configure each unit. The three systems were characterized and cross calibrated in the lab and during deployment in the experiment. These units were designed to operate primarily in the frequency domain for software radio. We found they operated satisfactorily in the time-domain required for our experiments if and only if the base-band frequency variable was set to 0. Otherwise, the gain is strongly dependent of this setting.

Code availability

Project specific computer code for data acquisition and analysis and for calculation of acceptances are not general-purpose codes and are, therefore, not user friendly. However, they are available upon reasonable request to the corresponding author.

**Data Availability**

All non-null data generated or analysed during this study are included in this published article and its Supplementary Information files. The raw data that were processed to find only background noise The raw datasets that were found to consist of only background signals and inconsistent with MQN impacts are not publicly available due to their large size and low expected interest but are available from the corresponding author on reasonable request.

**References**

[1] Bodmer, A. R. Collapsed nuclei, *Phys. Rev. D* **4,** 1601-1606 (1971). https://doi.org/10.1103/PhysRevD.4.1601

[2] Witten, E. Cosmic separation of phases, *Phys. Rev. D* **30,** 272-285 (1984). https://doi.org/10.1103/PhysRevD.30.272

[3] Farhi, E. and Jaffe, R. L. Strange matter, *Phys. Rev. D* **30,** 2379-2391 (1984). https://doi.org/10.1103/PhysRevD.30.2379

[4] Liang, X. and Zhitnitsky, A. Axion field and the quark nugget's formation at the QCD phase transition, *Phys. Rev. D* **94,** 083502 (2016). https://doi.org/10.1103/PhysRevD.94.083502

[5] Tatsumi, T. Ferromagnetism of quark liquid, *Phys. Lett. B* **489,** 280-286 (2000). https://doi.org/10.1016/S0370-2693(00)00927-8

[6] VanDevender, J. P., *et al.* Detection of magnetised quark-nuggets, a candidate for dark matter, *Sci. Rep.* **7,** 8758 (2017). https://doi.org/10.1038/s41598-017-09087-3

[7] Madsen, J. Physics and astrophysics of strange quark matter in *Hadrons in Dense Matter and Hadrosynthesis, Lecture Notes in Physics*, **516,** (eds. Cleymans J., Geyer H.B., Scholtz F.G.), 162-203 (Springer, 1999). https://www.springer.com/gp/book/9783662142387

[8] Cecchini, S., *et al.* Results of the search for strange quark matter and Q-balls with the SLIM experiment, *Eur. Phys. J. C* **57,** 525-533 (2008). https://doi.org/10.1140/epjc/s10052-008-0747-7

[9] Orito, S., *et al.* Search for supermassive relics with a 2000-m$^2$ array of plastic track detectors, *Phys. Rev. Lett* **66,** 1951-1954 (1991). https://doi.org/10.1103/PhysRevLett.66.1951

[10] Price, P. B. Limits on contribution of cosmic nuclearites to galactic dark matter, *Phys. Rev. D* **38,** 3813-3814 (1988). https://doi.org/10.1103/PhysRevD.38.3813

[11] The MACRO Collaboration, Ambrosio, M. *et al.* Nuclearite search with MACRO detector at Gran Sasso, *Eur. Phys. J. C* **13,** 453-458 (2000). https://doi.org/10.1007/s100520050708

[12] Porter, N.A., Fegan, D.J., MacNeill, G.C., and Weekes, T. C. A search for evidence for nuclearites in astrophysical pulse experiments, *Nature* **316,** 49 (1985). https://doi.org/10.1038/316049a0

[13] Piotrowski, L. W., *et al.* Limits on the flux of nuclearites and other heavy compact objects from the Pi of the Sky project, *Phys. Rev. Lett.* **125,** 091101 (2020). https://doi.org/10.1103/PhysRevLett.125.091101

[14] VanDevender, J. P., Shoemaker, I. M., Sloan, T., VanDevender, A.P., and Ulmen, B.A. Mass distribution of magnetised quark nugget dark matter and comparison with requirements and observations, *Sci. Rep.* **10,** 17903 (2020). https://doi.org/10.1038/s41598-020-74984-z

[15] De Rujula, A., and Glashow, S. Nuclearites – a novel form of cosmic radiation, *Nature* **312,** 734-737 (1984). https://doi.org/10.1038/312734a0

[16] Read, J. I. The local dark matter density, J. Phys. G: *Nucl. Part. Phys.* **41,** 063101 (2014). 10.1088/0954-3899/41/6/063101

[17] Westwood, E. K. A normal mode model for acousto-elastic ocean environments, *J. Acoust. Soc. Am.* **100,** 3631-3645 (1996). https://doi.org/10.1121/1.417226

[18] Dittman, Gerald. L. Calculation of brine properties, UCID Report 17406, LLNL, Livermore, CA USA. February 16, 1977. https://doi.org/10.2172/7111583

[19] Colman, Steven M., Kelts, Kerry R., and Dinter, David A. Depositional history and neotectonics in Great Salt Lake, Utah, from high-resolution seismic stratigraphy. *Sedimentary Geology* **148,** 61–78 (2002). https://digitalcommons.unl.edu/usgsstaffpub/277

[20] Finch, E. Strangelets: who is looking and how. *J. Phys G* **32,** S251 (2006). https://doi.org/10.1088/0954-3899/32/12/S31

[21] Burdin, S. *et al.* Non-collider searches for stable massive particles. *Phys. Rep.* **582,** 1 (2015). https://doi.org/10.1016/j.physrep.2015.03.004

[22] The CTA Consortium. *Science with the Cherenkov Telescope Array*, 1-338 (World Scientific, 2019). https://doi.org/10.1142/10986

[23] The JEM-EUSO Collaboration., Adams, J.H., Ahmad, S. *et al.* JEM-EUSO: Meteor and nuclearite observations. *Exp Astron* **40,** 253–279 (2015). https://doi.org/10.1007/s10686-014-9375-4

[24] VanDevender, J. P., *et al.* Results of search for magnetised quark-nugget dark matter from radial impacts on earth, *Universe* **7,** no. 5, 116 (2021). https://doi.org/10.3390/universe7050116

[25] VanDevender, J. P., VanDevender, Aaron P., Wilson, Peter, Hammel, Benjamin F., and McGinley, N. Limits on magnetized quark-nugget dark matter from episodic natural events, *Universe* **7,** no. 2, 35 (2021). https://doi.org/10.3390/universe7020035

[26] VanDevender, J. P., Buchenauer, C. J., Cai, C., VanDevender, A. P., and Ulmen, B. A. Radio frequency emissions from dark-matter-candidate magnetised quark nuggets interacting with matter, *Sci. Rep.* **10,** 13756 (2020). https://doi.org/10.1038/s41598-020-70718-3

**Acknowledgements**

We gratefully acknowledge S. V. Greene for first suggesting that quark nuggets might explain the geophysical evidence that initiated this research (she generously declined to be a coauthor); D. J. Fegan for especially helpful suggestions during this work; Benjamin Hammel for conducting laboratory experiments on the interaction of a simulated quark nugget with matter. The team is grateful to Harbor Master David Shearer of Utah State Parks, Ms. Jamie Phillips-Barnes of Utah State Lands, Mr. David Ghizzone and Mr. Chad Wadell of Gonzo Boat Rentals, Dr. Cory Angeroth of the US Geological Survey for permitting and technical assistance, and to Dr. Ed Atler, Mr. Karl Scheuch, Mr. Andrew Bloemendaal, Mr. Jonathon Cross, Mr. Red Atwood, Mr. James Bell, and Mr. Michael Rymer for technical support. This work was supported primarily by VanDevender Enterprises, LLC, with an encouraging contribution by Dr. Karl VanDevender. The CTH simulations were supported by the New Mexico Small Business Assistance Program through Sandia National Laboratories, a multi-program laboratory operated by Sandia Corporation, a wholly owned subsidiary of Lockheed Martin Company, for the U.S. Department of Energy's National Nuclear Security Administration under contract DE-AC04–94AL85000). By policy, work performed by Sandia National Laboratories for the private sector does not constitute endorsement of any commercial product. This work was supported by VanDevender Enterprises, LLC.

**Author Contributions**

T. S. was the project particle physicist. He analysed previously published limits on non-magnetized-quark-nugget dark matter from atmospheric observations and reinterpreted them to find limits on magnetized quark nuggets, calculated the acceptances for magnetized quark nuggets, analysed both calibration and observational data from the Great Salt Lake experiment for limits on both non-magnetized and magnetized quark nuggets, wrote the paper, and decided on the final wording.

J.P.V. was lead physicist and principal investigator. He planned and executed the experiment, conducted the explosive calibrations, took the data, did the coincidence analysis, contributed to the writing of the paper, prepared the figures, revised the paper to incorporate the improvements from the other authors, and concurred with the paper.

T. A. N. was the principal physicist for acoustics. She developed the acoustic waveguide model for the Great Salt Lake, developed and ran the final ORCA simulations, decided on the interpretation of the explosive calibration waveform, contributed to the writing and revision of the paper, and concurred with the paper.

R. L. B. measured the sound speed as a function of location and depth in the Great Salt Lake and concurred with the paper.

G. F. performed the initial ORCA simulations of acoustic propagation in the Great Salt Lake waveguide supervised by T. A. N., helped revise the paper, and concurred with the paper.

C. S. was the project software programmer and systems analyst. He developed, maintained, and operated the automated data collection and system monitoring software and developed the software for data management, verification, and analysis to upload data into the project database, filter and characterize the data for subsequent analysis, helped revise the paper, and concurred with the paper.

R. Z. was the project digital signal processing engineer. He did the initial configuration and adaption of the gnuradio software and Universal Software Radio Peripheral to meet project needs, helped revise the paper, and concurred with the paper.

H. J. developed data extraction and viewing software to permit real-time remote monitoring of the data through the bandwidth-limited data link in hostile environmental conditions. He concurred with the final paper.

**Additional Information**

Competing Financial Interests

The authors declare that there are no competing financial interests.

**Figure Captions**

**Figure 1.** Absolute rectified value of the pressure from calibration shots at a sensor a) at 3100 m and b) 4100 m distance from a 5-m length of PETN primer cord depositing 260 kJ/m energy/length. The cord extended vertically from the surface to the lake bottom.

**Figure 2.** Peak pressure at the sensor in the ~250 Hz oscillatory portion versus distance to the source for explosive calibrations using 130 kJ/m (blue) and 260 kJ/m (red) charges, each 5 meters long. The two horizontal lines give the pressure that will trigger recording 80% (dashed line) and 20% (dotted line) of the total observation time, as explained in the text and in Supplemental Information on Methods: Variation in trigger level from weather effects.

**Figure 3.** Pressure versus time for 3 longest duration signals recorded on (a) 11 May 2019, (b) 2 July 2019, and (c) 14 September 2019 for sensor a1 (blue), a2 (bright red) and a3 (dark red).

**Figure 4.** Flux limits as a function of MQN mass for 3 different surface magnetic fields $B_o$ of the MQN: a) 0.1 TT, b) 1.0 TT, and c) 3.2 TT. The blue (red) lines are for MQN density of $10^{18}$ ($3 \times 10^{17}$) kg/m$^3$. Two curves for each density show the observationally determined limits with zero streaming velocity (solid) and with a streaming velocity of 230 km/s through the Galaxy (dashed). The solid black line is the theoretically expected flux on the GSL assuming all the DM is made up of MQNs of a fixed mass. The black dotted line is the integrated flux greater than the mass on the abscissa theoretically predicted from the Aggregation Model (see text) for $B_o$=0.1 TT. Curves for higher $B_o$ have flux limits $< 10^{-17}$ m$^2$ s$^{-1}$ sr$^{-1}$ (do not fit on the plots) and are not constrained by the experimental flux limits. Values of $B_0 > 3.2$ TT were excluded as described in Ref. [14].

**Figure 5.** Plot of total MQN flux for mass greater than M as a function of $B_o$. The bottom of the grey triangle shows the flux limit for exclusion by the data from the GSL experiment with 90%

CL. The bottom of the blue triangle indicates exclusion by the telescope data from Piotrowski, *et al.* and Porter, *et al.* applied to MQNs.

**Tables**

None